\documentstyle[11pt,epsfig]{article}

\begin{document}
\begin{center}

{\large\bf Quantum Phase Transitions in the\\ $U(5)- O(6)$ Large $N$
limit} \vskip.4truecm
\vskip.4truecm
{\normalsize Feng Pan,$^{1,2}$ Yu Zhang,$^{1}$ and
J. P. Draayer$^{2}$}
\vskip .2cm {\small
$^{1}$Department of Physics, Liaoning Normal University, Dalian
116029, P. R. China\vskip .1cm
$^{2}$Department of Physics and Astronomy,
Louisiana State University, Baton Rouge, LA 70803-4001, USA}
\end{center}
\vskip.3cm

\noindent {\bf Abstract~}{\it
The $U(5)-O(6)$ transitional behavior of the Interacting Boson Model in
the large $N$ limit is revisited. Some low-lying energy levels, overlaps of
the ground state wavefunctions, $B(E2)$ transition rate for the decay of the first excited
energy level to the ground state, and the order parameters are
calculated for different total numbers of bosons. The results show that
critical behaviors of these quantities are greatly enhanced
with increasing of the total number of bosons $N$, especially
fractional occupation probability for $d$ bosons in the
ground state, the difference between the expectation value of
$n_d$ in the first excited $0^+$ state and the ground state,
and another quantity related to the isomer shift behave similarly
in both the $O(6)-U(5)$ large $N$
and $U(5)-SU(3)$ phase transitions.}
\vskip .3cm
\noindent {\bf Keywords}: Phase transitions, order parameters,
large $N$ limit.
\vskip .3cm
\noindent {\small\bf PACS numbers}: 21.60.Fw, 05.70.Fh, 21.10.Re, 27.70.+q
\vskip .5cm

Quantum phase transitions have been attracting a lot of attention in many
areas of physics. This is understandable because they are very important
for gaining a deeper understanding of various quantum many-body systems.$^{[1-3]}$
In atomic nuclei, such quantum phase transitions can be related to
different geometrical shapes of the
system, which can be described either by the Bohr-Mottelson model$^{[4]}$
(BMM) or by the Interacting Boson Model$^{[3]}$ (IBM). As summarized by
Iachello,$^{[5]}$ the study of shape phase transitions in atomic
nuclei was initiated in the early 80s$^{[6-8]}$ following some previous
work$^{[9]}$ by Gilmore.  It is now widely accepted that the three
limiting cases of the IBM correspond to three different geometric shapes
of nuclei, referred to as spherical (vibrational limit with $U(5)$
symmetry), axially deformed (rotational limit with $SU(3)$ symmetry),
and $\gamma$-soft (triaxial with $O(6)$ symmetry). This picture is
captured by the so-called Casten triangle.$^{[10]}$ More interesting
scenarios occur when a system is in between two different phases, in
which case a quantum phase transition occurs at the corresponding
critical point. A critical point at finite $N$ with $E(5)$ symmetry
along the $U(5)-O(6)$ leg of the Casten triangle was shown to exist in
[11], and many examples confirming the nature of this transition in
realistic nuclear system have been reported.$^{[12]}$ Recently this
transitional region has been studied for relatively large $N$ values and
the results show that the critical point region becomes progressively
narrower as the boson number $N$ increases.$^{[13,14]}$
This phenomenon has been explained in [14] in terms of a quasidynamical symmetry.
In order to study the large $N$ limit situation
corresponding to the classical BMM, one must approach the large $N$
limit from results for finite $N$ if algebraic results for the large $N$
limit is not available.  In this Letter we revisit the $U(5)-O(6)$ transitional
case in the large $N$ in detail to see whether there are substantial
changes that occur as $N$ grows ever larger, which serves as
a supplement to the results reported in [13] and [14].

Our investigation is based on the following schematic $U(5)-O(6)$
Hamiltonian:

$$H=(1-x)\hat{n}_{d}+{x\over{f(N)}}\hat{S}^+\hat{S}^-, \eqno(1)$$
where $\hat{n}_{d}=\sum\limits_{m}d_m^\dag d_m$ is the total number
of $d$-bosons, $ \hat{S}^+={1\over{2}}(d^\dag\cdot
d^\dag-s^{\dag2})$ and $\hat{S}^-=
{1\over{2}}(\tilde{d}\cdot\tilde{d}-s^2)$ are generalized boson
pair creation and annihilation operators,
$f(N)$ is a linear function of total number of bosons $N$, and
$x$ is the control parameter of the model.
It should be obvious that the system is in the $U(5)$ limit
when $x=0$ and in the $O(6)$ limit when $x=1$.
As the control parameter $x$ varies continuously within the
closed interval [0, 1], the system described by (1) undergoes a shape
(phase) transition from $U(5)$ to $O(6)$.

To diagonalize Hamiltonian (1), we expand the eigenstates of (1) in terms
of the $U(6)\supset U(5)\supset O(5)\supset O(3)$ basis vectors
$|N~n_{d}~v~LM\rangle$ as

$$|N~\xi~v~LM;~x\rangle=\sum\limits_{n_{d}}
C^\xi_{n_{d}}(x)|N~n_{d}~v~LM\rangle,\eqno(2)$$
where $C_{n_{d}}^\xi(x)$ is the expansion coefficient, $\xi$ is an
additional quantum number needed to label different eigenstates
with the same quantum numbers $v$, $L$, and, $M$.

To show how the energy levels change as a function of the control
parameter $x$ and the total number of bosons $N$, the lowest $25$ energy
levels as a function of $x$ for a system with fixed quantum number $\nu$
and $f(N)=N$ for $N=10,~40,~120$, and $300$ are shown in Fig. 1. It can
be seen from these results that there is a minimum in the low-lying
excitation energy when the control parameter has a value in the range
$0.45 < x < 0.65$, with the minimum growing sharper as the total
number of bosons increases. This control parameter region is
recognized as the critical point region of the vibrational-gamma soft
transition. To the left of the critical point region, $0 \leq x <0.45$,
there are $9$ degenerate levels ($x = 0$) that gradually split with
increasing $x$ into $25$ non-degenerate levels. Similarly, beyond the
critical region, $0.65 < x \leq 1$, the $25$ non-degenerate levels
coalesce into $5$ degenerate levels ($x = 1$). It should be noted
that apart from the end points, the levels are truly non-degenerate, and
that the level density grows rather dramatically within the critical
point region with increasing $N$; and furthermore, as $N$ grows the
critical point region becomes progressively narrower with a cusp around
$x\sim 0.45$, which is in agreement with the observation reported in
[14], in which only the $N=40$ case was shown.

\begin{figure}
\begin{center}
\epsfig{file=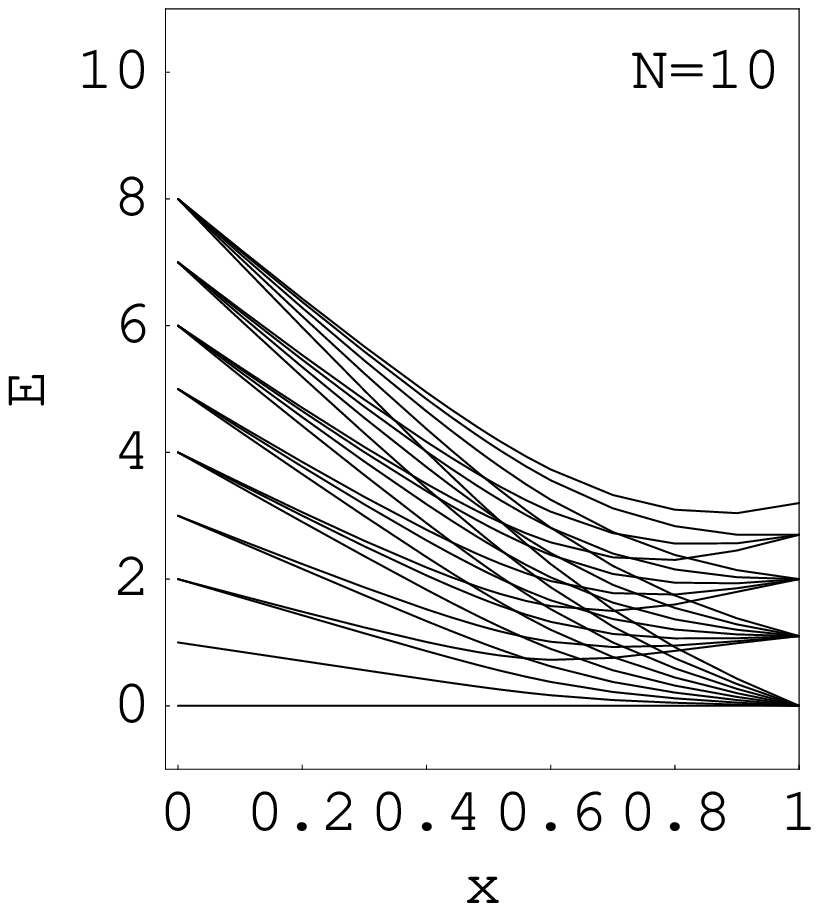,width=3.85cm}~~~
\epsfig{file=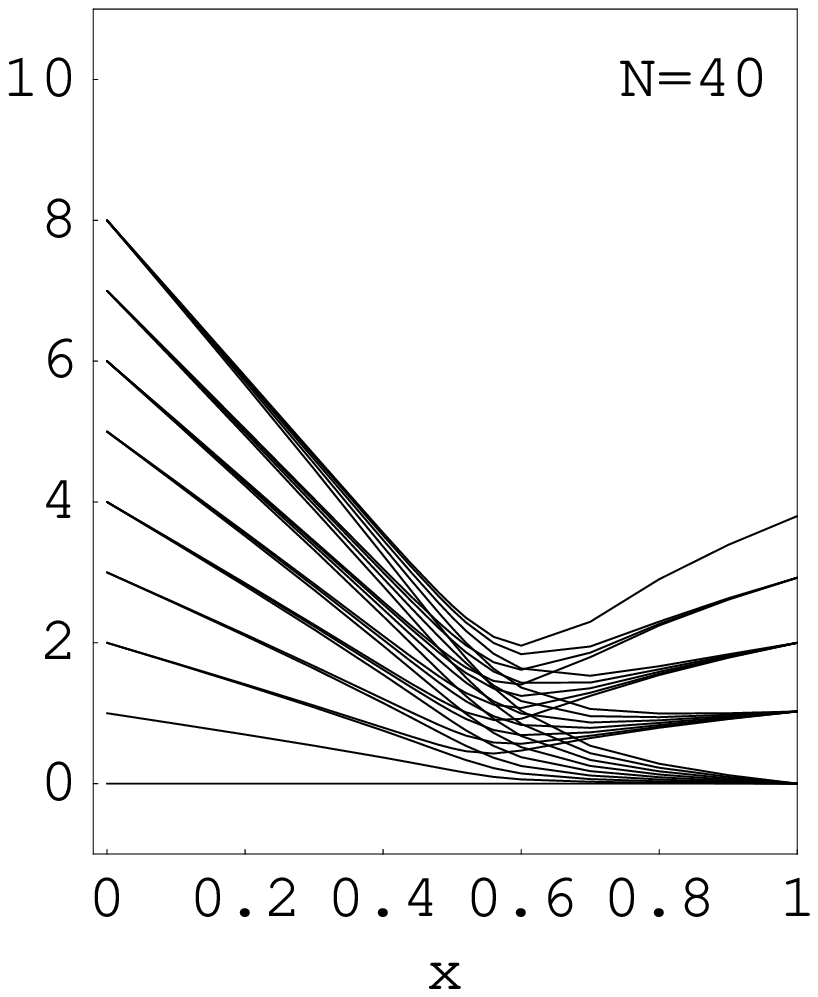,width=3.55cm}
\epsfig{file=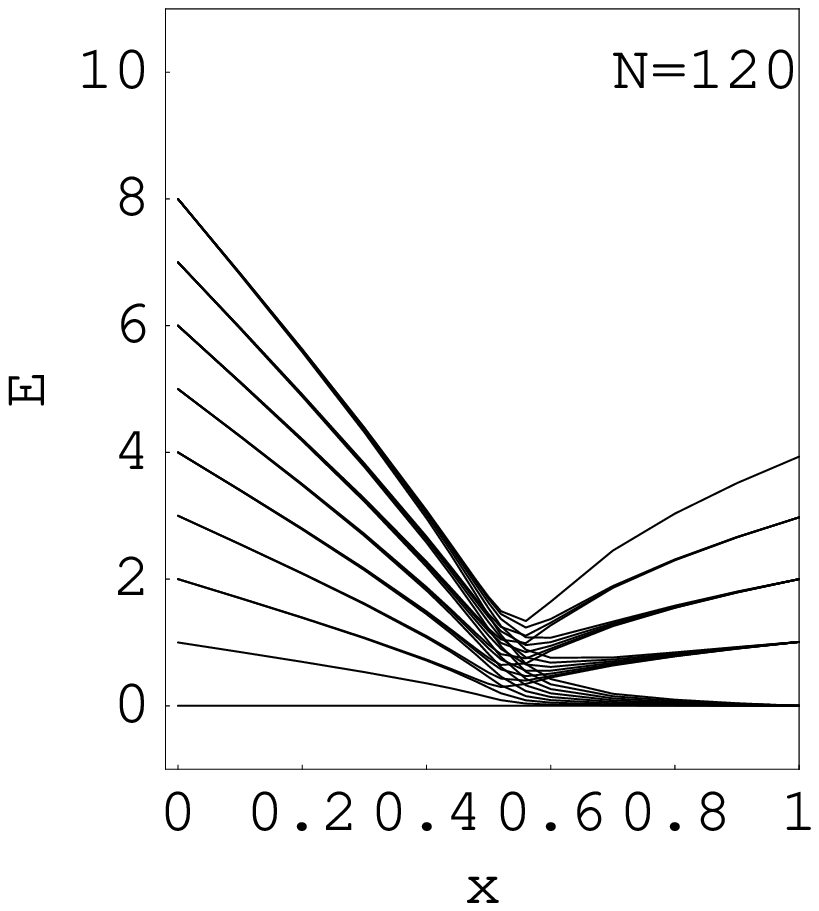,width=3.7cm}
\epsfig{file=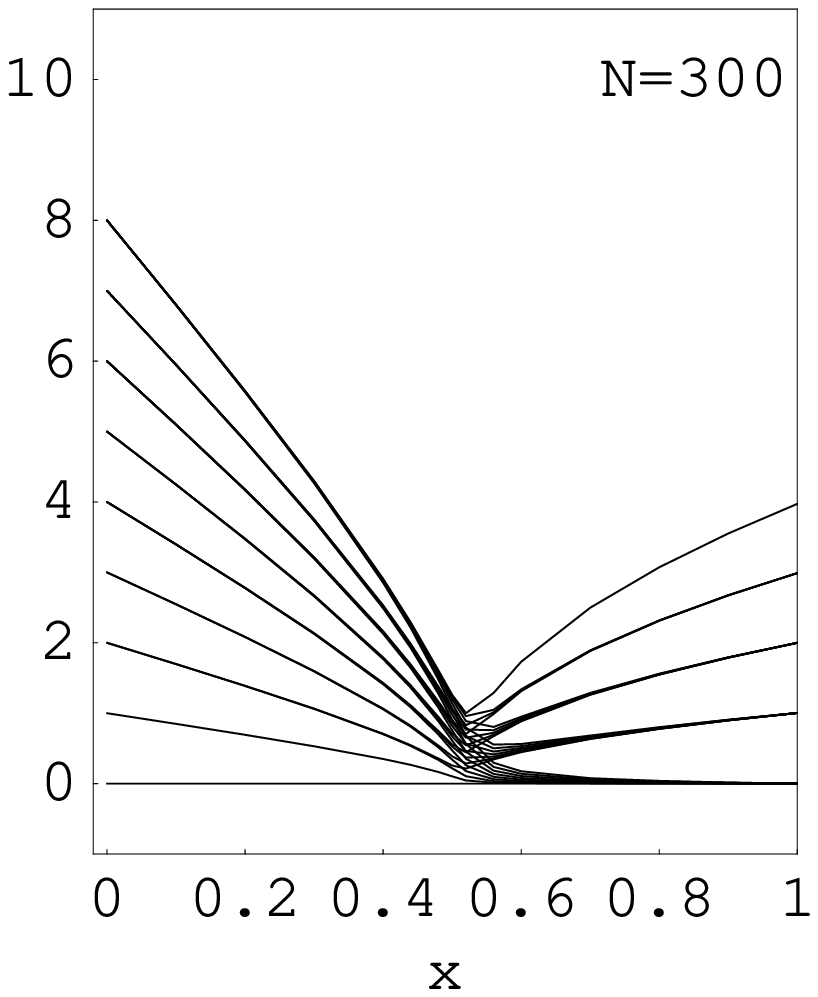,width=3.5cm}
\end{center}
{\scriptsize  ~Fig~1.~
The lowest $25$ energy levels (in arbitrary unit) of Hamiltonian (1)
with $f(N)=N$ as a function of $x$ for $N=10,~40,~120$, and $300$,
respectively.}
\end{figure}

\begin{figure}
\begin{center}
\epsfig{file=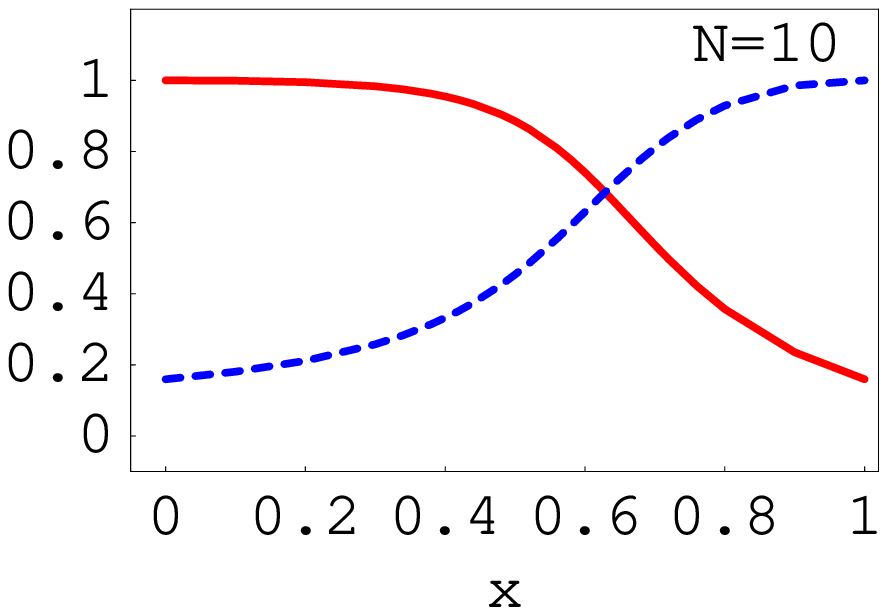,width=3.8cm}~~~
\epsfig{file=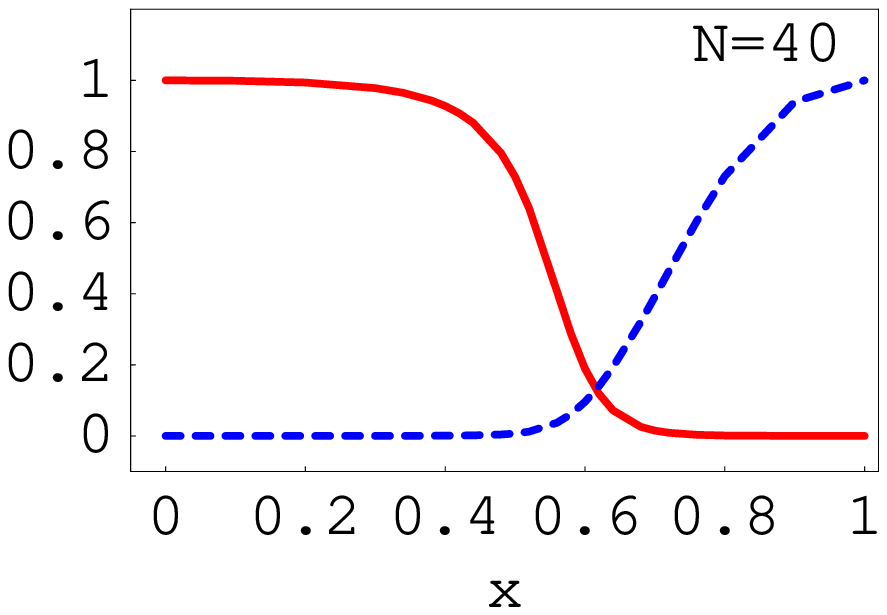,width=3.8cm}
\epsfig{file=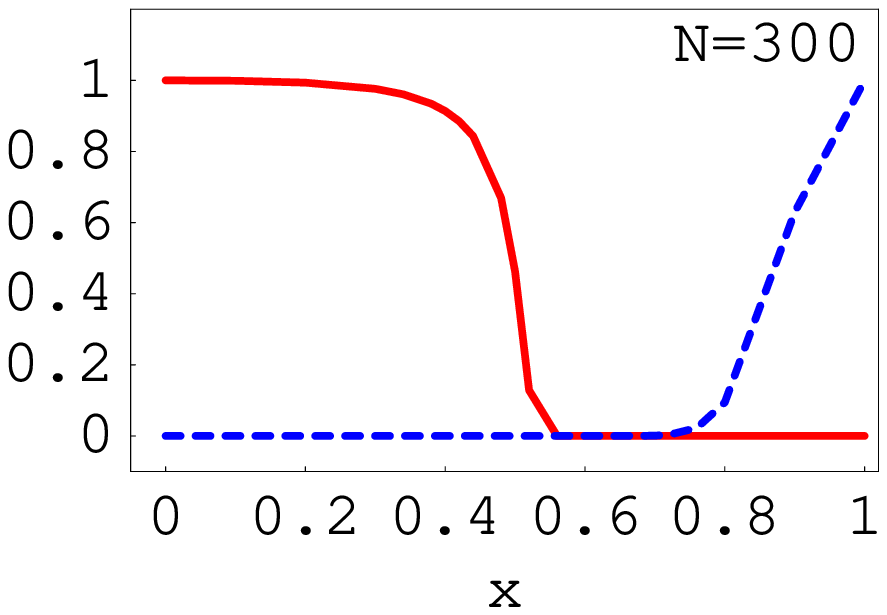,width=3.8cm}
\epsfig{file=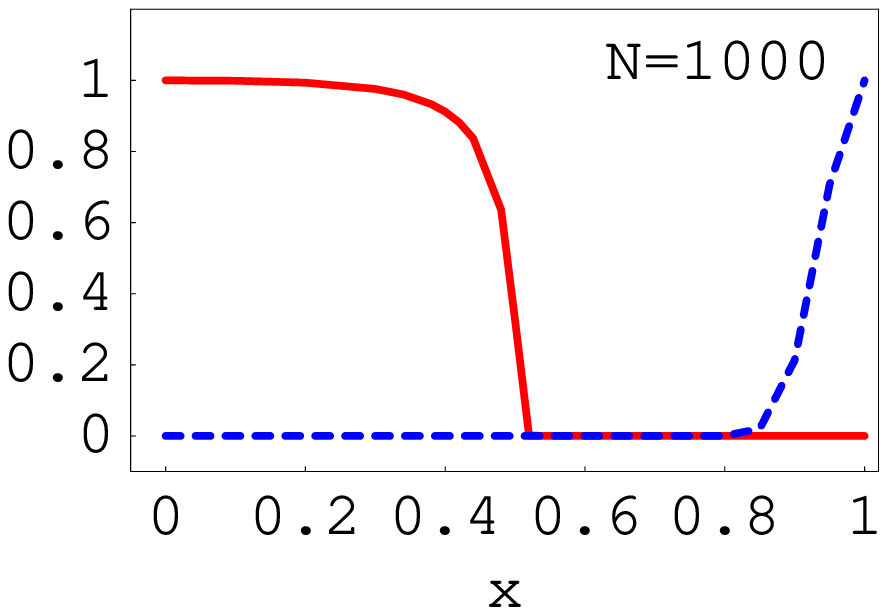,width=3.8cm}
\end{center}
{\scriptsize  Fig. 2. Overlaps of the ground state wavefunction, where
the full line shows the overlap $|\langle 0_g;x|0_g;x=0\rangle|$, and
the dotted line shows the overlap $|\langle 0_g;x|x_g;x=1\rangle|$.}
\end{figure}

The corresponding overlaps of the ground state wavefunctions of
Hamiltonian (1) as a function of the control parameter $x$ with those of
limiting cases $|\langle 0_g;x|0_g;x_0\rangle|$ for $x_0=0,~1$ for
different total number of bosons $N=10,~40,~300,~1000$ were also
calculated and the results are shown in Fig. 2.
It can be seen from Fig. 2 that there is a
cross-over point at a certain nonzero amplitude around $x\sim 0.65$
for the overlaps $|\langle 0_g;x|0_g;x=0\rangle|$
and $|\langle 0_g;x|x_g;x=1\rangle|$ when
$N$ is relatively small, which yields to a cross-over region
with near zero amplitude when $N$ becomes larger.
Furthermore, there is a sharp change in
$|\langle 0_g;x|0_g;x=0\rangle|$
around a critical point $x_{c}\sim 0.45$
in the large $N$ limit. These results suggest that the largest
absolute value of the derivative of
$|\langle 0_g;x|0_g;x=0\rangle|$
with respect to $x$ occurs around the critical
point in the large $N$ limit. While both
$|\langle 0_g;x|0_g;x=0\rangle|$ and
$|\langle 0_g;x|0_g;x=1\rangle|$ are all
rather smooth in the relatively small $N$ cases.

$B(E2)$ transition rates for decay of the first excited
$\nu=1$ energy level to the ground state for $N=10,~40,~300$, and
$1000$ expressed in units with $B(E2; 1\rightarrow 0)=100$ in the
$U(5)$ ($x=0$) limit were also calculated.  The $E2$ transition
operator was chosen as
$T(E2)=e_2{(s^\dag\tilde{d}+d^\dag\tilde{s})^2_q}$, where
$e_2$ is the effective charge. The results are shown in Fig. 3.
It can be seen quite clearly that the $B(E2; 1\rightarrow 0)$ changes
rather smoothly with $x$ for small $N$, while there is a
sharp change at the critical point when $N$ is large enough.
This behavior of the $B(E2; 1\rightarrow 0)$
was also reported for the $N\leq 60$ cases considered
in [14].

\begin{figure}
\begin{center}
\epsfig{file=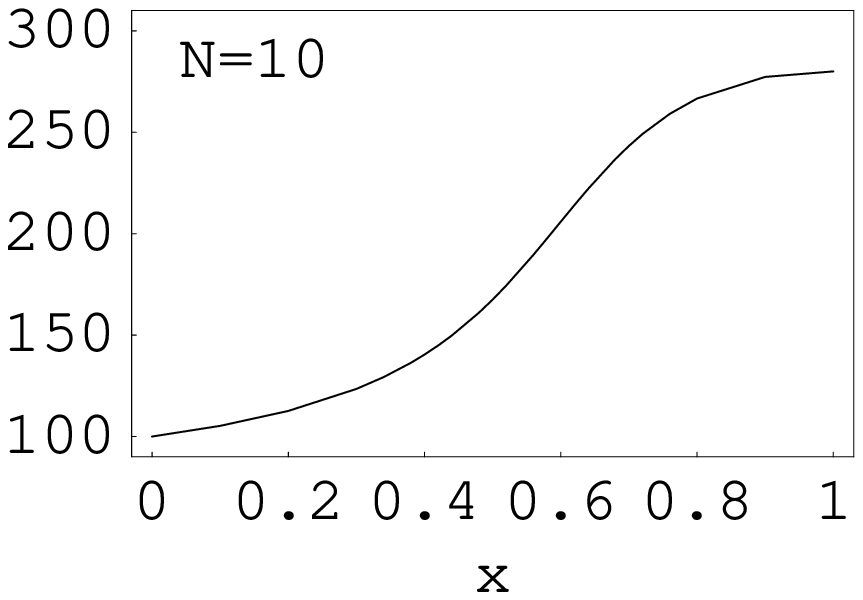,width=3.8cm}~~~
\epsfig{file=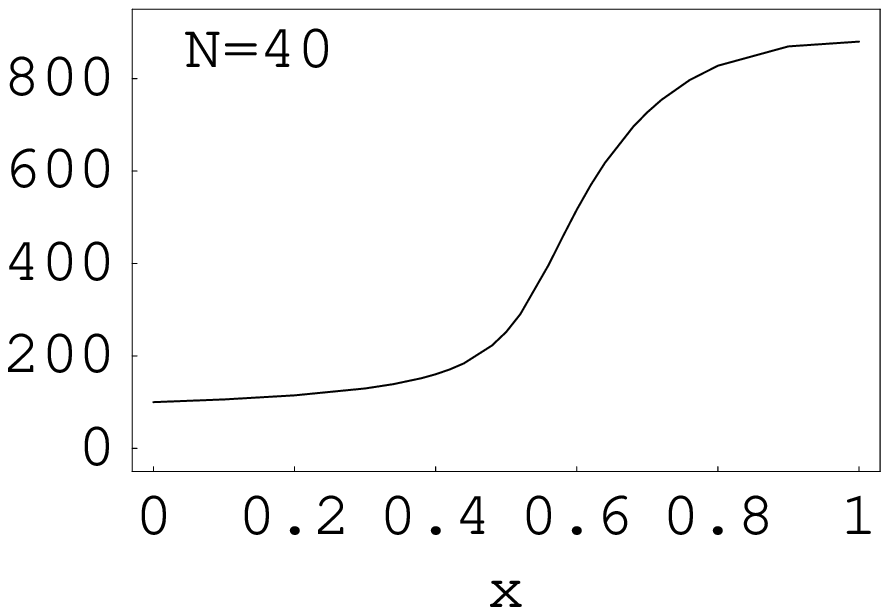,width=3.8cm}
\epsfig{file=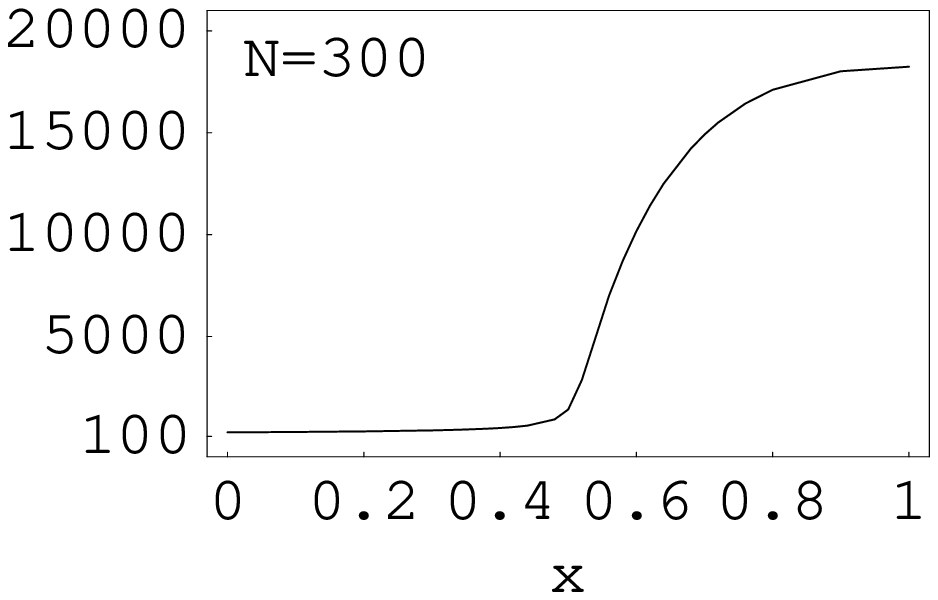,width=3.8cm}
\epsfig{file=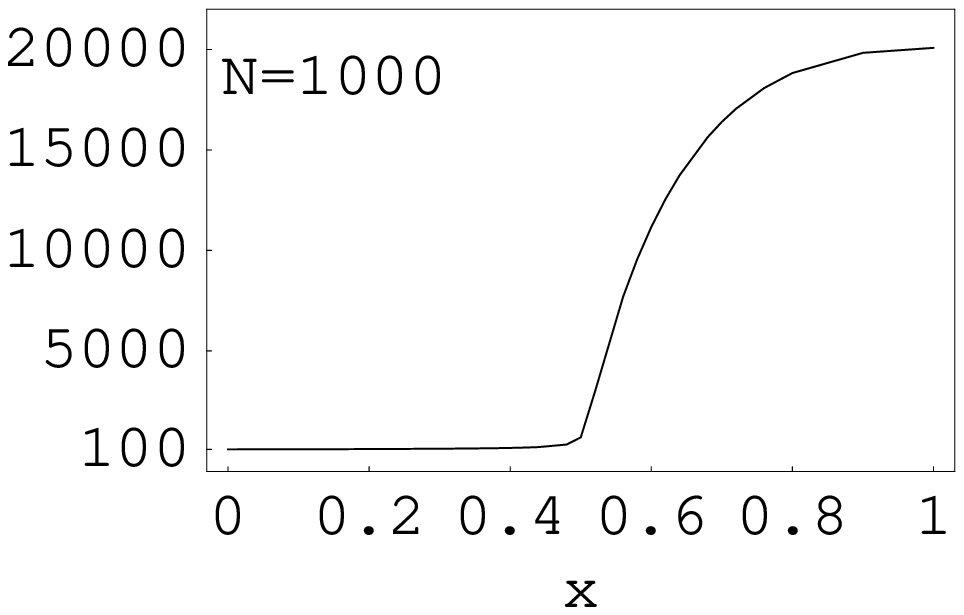,width=3.8cm}
\end{center}
{\scriptsize  Fig. 3.
$B(E2)$ transition rates for decay of the first excited
$\nu=1$ energy level to the ground state for $N=10,~40,~300$, and
$1000$ expressed in units with $B(E2; 1\rightarrow 0)=100$ in the
$U(5)$ ($x=0$) limit.}\end{figure}

The fractional occupation probability for $d$ bosons in the
ground state, $\rho_d=\langle\hat{n_d}\rangle/N$
as a function of $x$ was reported in [13] and [15].
It was shown that an order parameter to
signify a second-order phase transition can be
chosen to be $\rho_{d}$.
Our calculation indicates
that the system is almost in the $U(5)$ limit when
$x\sim0-0.45$ in the large $N$ limit, which
corresponds to an $s$-boson condensate.
The occupation probability $\rho_{d}$
gradually increases within the critical
region for relatively small $N$ values with the
change in  $\rho_{d}$ becoming
sharper and sharper with increasing $N$,
which is in agreement to the results
reported in [13] and [15].
Since the behavior of the order parameter $\rho_{d}$ is the same for
both first- and second-order transitions for the small $N$ cases,
in order to distinguish whether
the phase transition is of first or second order from model calculations,
another order parameter, the difference between the expectation value of
$n_d$ in the first excited $0^+$ state and the ground state, ${\rm
v}_{1}=\alpha_{0}(\langle 0_{2}\vert n_{d}\vert 0_{2}\rangle- \langle
0_{g}\vert n_{d}\vert 0_{g}\rangle)$, was introduced in [15]. The authors
showed that ${\rm v}_{1}$ displays a wiggling, sign-change-behavior in
the region of the critical point due to the switching of the two
coexisting phases, which is characteristic of a first-order transition,
while ${\rm v}_{1}$ has a smoother behavior that is characteristic of a
second-order transition. It should be pointed out that the conclusion
made in [15] are for finite $N$ only. However, since order of
a phase transition should always be defined in the thermodynamic limit,
an effective order parameter must also behave differently in phase
transitions with different orders.
To see whether the order parameter ${\rm v}_1$ and
another quantity ${\rm v}_2=
\beta_0(\langle 2_1|n_d|2_1 \rangle-\langle 0_1|n_d|0_1
\rangle)$ related to the isomer shift $\delta\langle r^2\rangle=
\langle r^2\rangle_{2_{1}}-\langle r^2\rangle_{0_g}$
introduced in [15] satisfy this criterion,
both ${\rm v}_1$ and ${\rm v}_2$
were calculated for the $N=10,~40,~300,$ and $1000$ cases. The results
are shown in Fig. 4. In order to compare curves of
${\rm v}_{1}$ and ${\rm v}_{2}$ for the different $N$ cases, the
parameters $\alpha_{0}$ and $\beta_{0}$ were taken to be $1$ instead of
the $1/N$ used in [15]. Our calculation shows that: (a) both ${\rm v}_1$
and ${\rm v}_{2}$ have a smooth behavior when $N$ is relatively small; (b)
${\rm v}_{1}$ gradually displays of a sign-changing nature in the critical
region when $N$ is large enough, with this behavior being greatly enhanced
in the large $N$ limit; and (c) there is an obvious peak in ${\rm v}_2$
in the large $N$ limit, while ${\rm v}_2$ is rather smooth for relatively
small $N$. These results shown  in Fig. 4, together with those shown in
[15], indicate that the order parameters ${\rm v}_1$ and
${\rm v}_2$, like another order
parameter $\rho_{d}$, behave similarly in both
the $O(6)-U(5)$ large $N$ and $U(5)-SU(3)$ phase transitions.
Due to current computation limitation, one can not
calculate these quantities in the
$U(5)-SU(3)$ transitional case exactly for $N\geq 30$.
Therefore, whether these quantities
in the $U(5)-SU(3)$ case will change
substantially in the large $N$ limit is still
an open question.
For relatively small $N$ cases, however,
as indicated in [15], the order parameters ${\rm v}_1$ and ${\rm v}_2$
are indeed behave differently in the $O(6)-U(5)$ and
$U(5)-SU(3)$ phase transitions, which, therefore,
can be used to signify the order of the transition
from the small $N$ cases.

\begin{figure}
\begin{center}
\epsfig{file=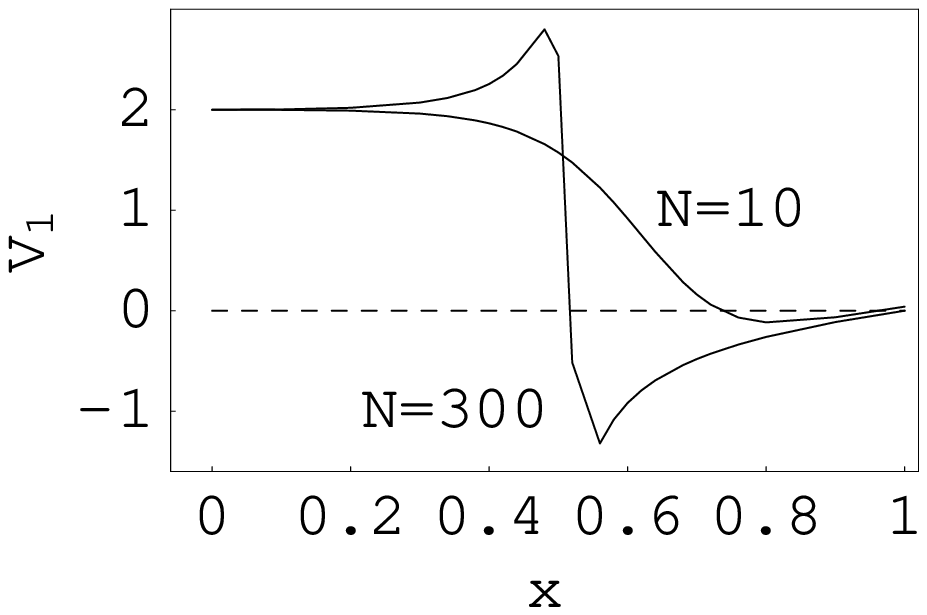,width=3.8cm}~~~
\epsfig{file=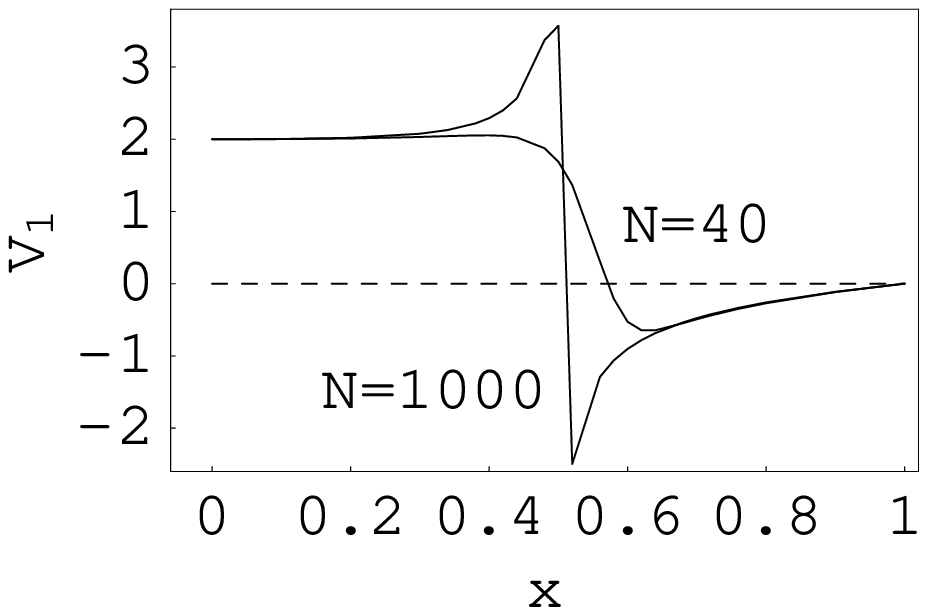,width=3.8cm}
\epsfig{file=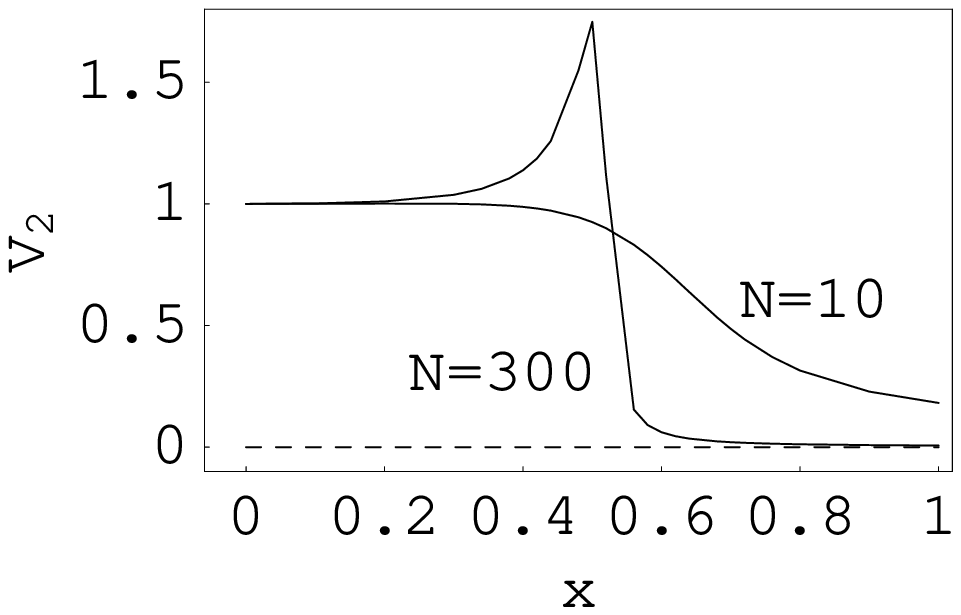,width=3.8cm}
\epsfig{file=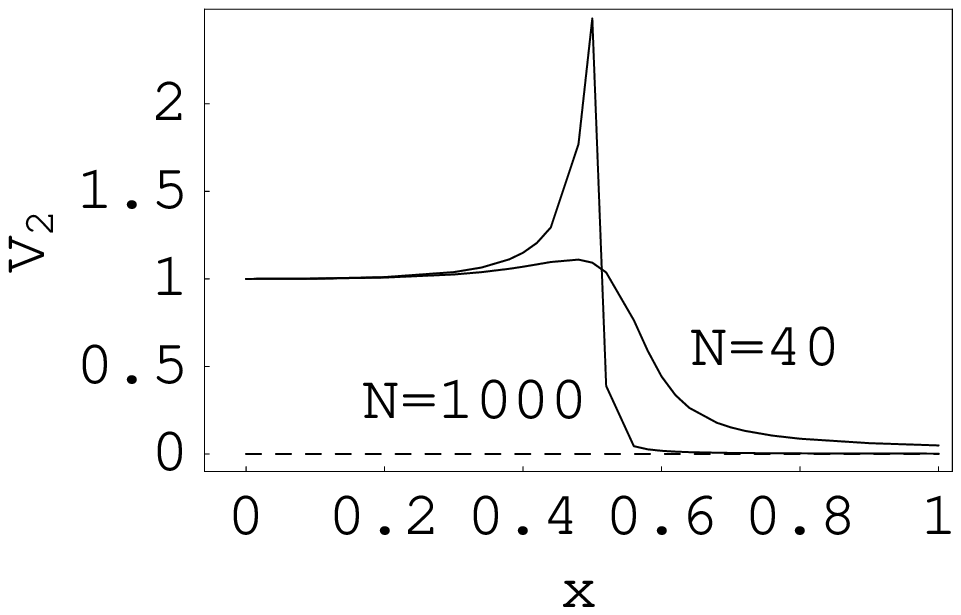,width=3.8cm}
\end{center}
{\scriptsize  Fig. 4. Behavior of the order parameters ${\rm v}_1$
and ${\rm v}_2$ as functions of the control parameter $x$ for different
$N$ values, where the parameters $\alpha_0$ and $\beta_0$ in
${\rm v}_1$ and ${\rm v}_2$, respectively, are set to be $1$.}
\end{figure}

In summary, the $O(6)-U(5)$ transitional behavior in the large
$N$ limit has been revisited. Some low-lying energy levels, overlaps of
the ground state wavefunctions,
$B(E2)$ transition rates for decay of the first excited $\nu=1$ energy level to
the ground state, and the order parameters
${\rm v}_1$ and ${\rm v}_2$ related to
the isomer shifts were calculated for different total number of bosons.
It is found that the critical behaviors of these quantities are
greatly enhanced with increasing of the total number of bosons $N$,
especially all the order parameters, $\rho_d$, ${\rm v}_1$, and ${\rm v}_2$
behave similarly in both the $O(6)-U(5)$ large $N$
and $U(5)-SU(3)$ phase transitions.
The drastic enhancement of these quantities near the critical point may
be explained in terms of a quasidynamical symmetry.$^{[14]}$
The ``specific heat'' introduced in [16]
seems also suitable to be used to
classify the order of the phase transitions since
these quantities behave quite differently
in first and second order phase transitions
even when the $N$ is finite.

In the IBM for atomic nuclei the total number of bosons $N$ is
phenomelogically related to be the number of valence  $s$ and $d$ nucleon
pairs, which is usually a relatively small number.
However, in the large $N$ limit, the IBM
yields to the BMM, in which there is no restriction on the number
of bosons; indeed, in principle, this should correspond to the $N
\rightarrow \infty$ limit.
Therefore, the results shown in this Letter should be helpful
in understanding the nature of the vibration to gamma-soft phase
transition in the BMM. It is
interesting to check to see whether there are substantial differences
between the $E(5)$ symmetry derived from an extreme case of the BMM and
systems described by a $U(5)-O(6)$ Hamiltonian with the finite $N$
based on the IBM. A recent
study suggest that the  $E(5)$ symmetry can only be described
approximately in the IBM,$^{[17]}$ which is a conclusion that is
consistent with our results.

\vskip .2cm
Support from the U.S. National Science Foundation
(0140300), the Southeastern Universitites Research
Association, the Natural Science Foundation of China
(10175031), the Education Department of Liaoning Province,
and the LSU-LNNU joint research program (C164063)) is acknowledged.

\def\HT{\bf\relax}
\def\REF#1{\small\par\hangindent\parindent\indent\llap{#1\enspace}\ignorespaces}

\section*{References}

\noindent\REF{[1]}S. Sachdev, {\it Quantum Phase Transitions} (Cambridge
Univ. Press, Cambridge, 1999).

\REF{[2]} J. A. Hertz, Phys. Rev. B {\bf 14}, 1165 (1976).

\REF{[3]} F. lachello and A. Arima, {\it The Interacting Boson Model}
(Cambridge University
Press, Cambridge, 1987).

\REF{[4]} A. Borh and  B. R. Mottelson, {\it Nuclear Structure} Vol. I
(Benjamin, New York, 1969); Vol. II (Benjamin, New York, 1975).

\REF{[5]} F. lachello, AIP Conf. Proc. {\bf 726},
111 (2004).

\REF{[6]} A. E. L. Dieperink, O. Scholten, and F. lachello, Phys.
Rev. Lett. {\bf 44}, 1747 (1980).

\REF{[7]} D. H. Feng, R. Gilmore, and S. R. Deans, Phys. Rev. {\bf
C23}, 1254 (1981).

\REF{[8]} O. S. Van Roosmalen, {\it Algebraic Description of Nuclear
and Molecular
Rotation-Vibration Spectra}, Ph.D. Thesis, University of Groningen,
The Netherlands, 1982.

\REF{[9]} R. Gilmore and D. H. Feng, Nucl. Phys. {\bf A301}, 189 (1978);
R. Gilmore, J. Math. Phys. {\bf 20}, 89 (1979).

\REF{[10]} R. F. Casten,  in {\it Interacting Bose-Fermi System},
ed. F. Iachello (Plenum, 1981).

\REF{[11]} F. Iachello, Phys. Rev. Lett. {\bf 85}, 3580 (2000).

\REF{[12]} R. M. Clark, M. Cromaz, M. A. Deleplanque, M. Descovich,
R. M. Diamond, P. Fallon, I. Y. Lee, A. O. Macchiavelli,
H. Mahmud, E. Rodriguez-Vieitez, F. S. Stephens, and D. Ward,
Phys. Rev. {\bf C69}, 064322 (2004).

\REF{[13]} J. M. Arias, J. Dukelsky, and J. E. Garcia-Ramos,
Phys. Rev. Lett. {\bf 91}, 162502 (2003).

\REF{[14]} D. J. Rowe, Phys. Rev. Lett. {\bf 93}, 122502 (2004).

\REF{[15]} F. Iachello, N. V. Zamfir, Phys. Rev. Lett. {\bf 92}, 212501 (2004).

\REF{[16]} P. Cejnar, S. Heinze, and J. Dobe{\v{s}},
Phys. Rev. C {\bf 71}, 011304 (2005).

\REF{[17]} J. M. Arias, C. E. Alonso, A. Vitturi, J. E. Garcia-Ramos,
J. Dukelsky, A. Frank, Phys. Rev. {\bf C68}, 041302 (2003).
\end{document}